\documentclass[twocolumn,aps,prc,preprintnumbers,showpacs,showkeys,nofootinbib,superscriptaddress,fleqn,floatfix,tightenlines, 10pt]{revtex4-1}
\usepackage{amsmath,amsfonts,amssymb,amscd,amsxtra,amsthm}
\usepackage{natbib}
\usepackage{graphicx,subfigure}  
\usepackage{epstopdf}
\usepackage{dcolumn}  
\usepackage{bm}          
\usepackage{slashed}
\usepackage[utf8]{inputenc}
\usepackage{CJK}
\usepackage{mathptmx}
\usepackage{mathrsfs}
\usepackage[normalem]{ulem} 
\usepackage[dvipsnames]{xcolor} 
\usepackage{array}
\usepackage{multirow}
\renewcommand\arraystretch{1.6}

\begin{document}

\title{Short-range correlations and momentum distributions in mirror nuclei $\rm ^3H$ and $\rm ^3He$}
\author{Qi Meng}
\address{School of Physics, Nanjing University, Nanjing 210093, China}
\author{Ziyang Lu}
\address{School of Physics, Nanjing University, Nanjing 210093, China}
\author{Chang Xu}
\email{cxu@nju.edu.cn}
\address{School of Physics, Nanjing University, Nanjing 210093, China}

\begin{abstract}
Motivated by recent high-energy electron and $\rm ^3H$ and $\rm ^3He$ nuclei scattering experiment in Jefferson Lab (Nature 609, 41 (2022)), 
the short-range correlations (SRCs) between nucleon pairs for 3-nucleon systems are microscopically studied using realistic $NN$ 2-body interaction and two-Gaussian type $NNN$ 3-body interaction.
The wave functions of both $\rm ^3H$ and $\rm ^3He$ are obtained by solving 3-body Schr\"{o}dinger equations using Gaussian expansion method (GEM).
The differences of one-nucleon and nucleon-nucleon momentum distributions between $\rm ^3H$ and $\rm ^3He$ are analyzed in detail. The results show that the percentages of $pn$-SRC pairs are significantly enhanced as compared with those of $nn(pp)$-SRC ones in $\rm ^3H$ and $\rm ^3He$ nuclei, which is consistent with the experimental findings.
\end{abstract}

\maketitle

\section{Introduction}

Short-range correlations (SRCs) between pairs of nucleons are important aspects in nuclear physics, which are considered to be generated from the strong, short-distance part of nucleon-nucleon ($NN$) interactions. SRCs are important for comprehensive understanding of not only the essential feature of nuclear dynamics but also the nuclear forces at short distance and how they are generated from the strong interaction between quarks in nucleons \cite{Hen:2016kwk}. The nucleon-nucleon SRC pair is considered to have large relative momentum and small total momentum, leading to a high-momentum tail in one-nucleon and nucleon-nucleon momentum distributions. The study of SRCs and its high-momentum feature will deepen our understanding of the properties of finite nuclei at normal density and nuclear matter at supra-saturation density, which probably has important implications in determining the internal structure and evolution of stellar objects such as neutron stars.

Sophisticated theoretical approaches, using modern realistic interactions \cite{Reid:1968sq, Lacombe:1980dr, Pudliner:1997ck, Wiringa:1984tg, Wiringa:1994wb}, can be applied to study the correlated many-body wave functions and SRCs, such as correlated basis function theory \cite{Co:1994bzw, Fabrocini:1999mz, Bisconti:2007vu, Ryckebusch:2014ann}, self-consistent Green's function method \cite{Dickhoff:2004xx, Rios:2013zqa}, approximate schemes like cluster expansions \cite{Alvioli:2007zz, Alvioli:2012qa, Alvioli:2011aa, Alvioli:2016wwp}, Tensor-optimized high-momentum antisymmetrized molecular dynamics \cite{Lyu:2019bxr}, generalized nuclear contact formalism \cite{Weiss:2016obx} and variational Monte Carlo calculations \cite{Schiavilla:2006xx, Wiringa:2008dn, Wiringa:2013ala, Carlson:2014vla, Piarulli:2022ulk}. In general, the high-momentum tail of $pn$-SRCs in light nuclei have been demonstrated to be a universal feature with these state-of-art approaches. Various experimental efforts have also been devoted to the investigation of SRCs with the aim of probing the short-range properties of nuclear force \cite{Tang:2002ww, CLAS:2005ola, Piasetzky:2006ai, JeffersonLabHallA:2007lly, Subedi:2008zz, CLAS:2010yvl, Fomin:2011ng, LabHallA:2014wqo, Hen:2014nza, CLAS:2018xvc, CLAS:2018yvt}. Thanks to the high-energy and large momentum transfer electron and proton scattering experiments, it becomes possible to resolve the structure and dynamics of individual nucleons and nucleon pairs with precise measurements of small cross sections. Experimental data have showed that about 20$\%$ of the nucleons in nuclei have momentum larger than the Fermi momentum $k_F$ in saturated nuclear matter \cite{Hen:2016kwk,CiofidegliAtti:2015lcu, Frankfurt:1993sp, Frankfurt:2008zv, CLAS:2005ola, Arrington:2011xs}. Follow-up experiments probing the isospin composition of nucleon-nucleon SRCs were successfully conducted in both balanced and imbalanced nuclei, indicating that the $pn$-SRCs are much more dominating than the $pp$ and $nn$ ones. 

Recently, an experiment conducted in the Jefferson Lab accurately measured the $pn$-SRC pairs and $pp$-SRC ones in 3-nucleon systems, using high-energy electron and $\rm ^3H$ and $\rm ^3He$ nuclei scattering experiment \cite{Li:2022fhh}. 
This experiment took advantage of the mirror properties of $\rm ^3H$ and $\rm ^3He$ and avoided the direct measurement of high-momentum nucleons in the final state \cite{Li:2022fhh}, which improved the experimental accuracy and greatly reduced the uncertainties. Very interestingly, the experimental data show the ratio of $pn$-SRCs to $pp$-SRCs over the pair-counting prediction $P_{np/pp}=NZ/(Z(Z-1)/2)$ for $A=3$ nuclei is $2.17^{+0.25}_{-0.20}$, which is much smaller than that in heavy nuclei.

Motivated by this unexpected experimental result, we investigate the $pn$-SRCs and $pp$($nn$)-SRCs in mirror nuclei $\rm ^3H$ and $\rm ^3He$. We obtain both one-nucleon and nucleon-nucleon momentum distributions from an $ab\ initio$ calculation of solving 3-body Schr\"{o}dinger equation with a realistic $NN$ 2-body interaction, {\it i.e.} Argonne $v_8'$ (AV8') interaction, and a two-Gaussian type $NNN$ 3-body interaction.
The numerical method we applied to obtain the accurate correlated wave functions is the Gaussian expansion method (GEM) \cite{Hiyama:2003cu}, which has been successfully used in both nuclear physics and hadron physics \cite{Hiyama:2022jqh, Hiyama:2022loc, Hiyama:2018ukv}. For instance, we have applied the GEM to both bound and resonant states problems of tetraquark and pentaquark and satisfactory results have been obtained \cite{Meng:2019fan, Meng:2020knc, Meng:2021yjr}.
Realistic momentum distributions are obtained from the Fourier transform of correlated wave functions and the differences between SRCs in $\rm ^3H$ and $\rm ^3He$ are analyzed in detail in the present work. The comparison of SRCs in such imbalanced mirror nuclei with fully microscopic calculations may shed light on the equation of state (EoS) of asymmetric nuclear matter and the density-dependence of nuclear symmetry energy \cite{Xu:2010xh,Xu:2009bb,Xu:2012hf,Zhang:2014bna,Carbone:2011wk,Vidana:2011ap,Li:2019xxz}.  

\section{Methodology}

The Hamiltonian for a 3-nucleon system is given by
\begin{eqnarray}
H=\sum_{i=1}^3 \Big(m_i+ \frac{\boldsymbol{p}_i^2}{2m_i} \Big)-T_G+\sum_{i<j=1}^3 V_{ij}+V^{NNN},
\end{eqnarray}
where $m_i$ and $\boldsymbol{p}_i$ are the mass and momentum of the $i$-th nucleon, respectively. $T_G$ is the kinetic energy of the center-of-mass (c.o.m.) motion of the 3-nucleon system.
The complete 2-body interaction for a given $NN$ pair $ij$, $V_{ij}$, is composed of the strong interaction $V_{ij}^{NN}$ and electromagnetic interaction $V_{ij}^{EM}$,
\begin{eqnarray}
V_{ij}=V_{ij}^{NN}+V_{ij}^{EM}.
\end{eqnarray}
For the $NN$ strong interaction, we employ the Argonne $v_8'$ (AV8') interaction \cite{Pudliner:1997ck}. For the electromagnetic interaction, we consider the Coulomb force between proton-proton.
The 3-body interaction $V^{NNN}$ we applied is a two-Gaussian type $NNN$ 3-body interaction taken from Ref.\cite{Hiyama:2004nf}, which is optimized by fitting the bound states of $\rm{^3H}$, $\rm{^3He}$ and $\rm{^4He}$. Its function form is given by
\begin{eqnarray}
V^{NNN}=\sum_{n=1}^2 V_n^{(3)} e^{-\mu_n(r_{12}^2+r_{23}^2+r_{31}^2)}.
\end{eqnarray}

\begin{figure}[!htbp]
\centering
\includegraphics[width=8.6cm,angle=0,clip=true]{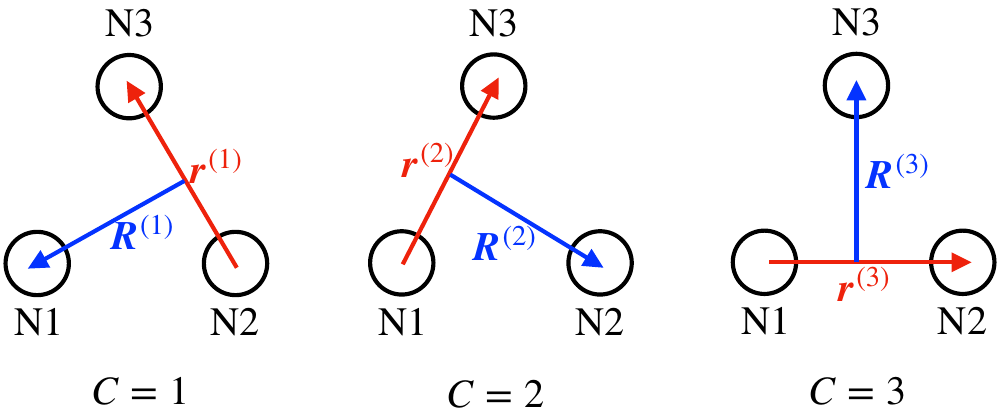}\\
  \caption{3-body Jacobi coordinates of 3-nucleon system in coordinate space.}
  \label{jacobi}
\end{figure}
In Gaussian expansion method, the variational wave function of a 3-nucleon system in coordinate space, $\Psi_{TM_T,JM}$ with isospin $T$, its $z$-component $M_T$, total angular momentum $J$ and its $z$-component $M$, is given by,
\begin{eqnarray}
&&\Psi_{TM_T,JM}=\sum_{C=1}^3 \sum_{\alpha}A_{\alpha}
\Big[ [\eta_{\frac{1}{2}}\eta_{\frac{1}{2}}]_t \eta_{\frac{1}{2}} \Big]_{TM_T} \times \notag \\
&& \ \ \ \ \ \ \ \ \Big[ \big[ [\chi_{\frac{1}{2}}\chi_{\frac{1}{2}}]_s \chi_{\frac{1}{2}} \big]_{S} \big[ \phi_{nl}(\boldsymbol{r}^{(C)} )\psi_{NL}(\boldsymbol{R}^{(C)}) \big]_I \Big]_{JM},
\end{eqnarray}
where $\eta_{\frac{1}{2}}$, $\chi_{\frac{1}{2}}$ are the isospin and spin wave functions of a single nucleon, respectively. $\phi$ and $\psi$ denote spatial wave functions with principal quantum number $n$, $N$ and orbit angular momentum $l$, $L$, respectively.


The label $(C)$ specifies a set of Jacobi coordinates shown in Fig.\ref{jacobi}. $A_{\alpha}$ specifies the expansion coefficients which are determined by matrix diagonalization, where the label $\alpha$ includes all quantum numbers for the expansion, $\alpha\equiv \{t,T,s,S,n,l,N,L,I \}$. $(J,T,M_T)$ are $(\frac{1}{2},\frac{1}{2},-\frac{1}{2})$ and $(\frac{1}{2},\frac{1}{2},\frac{1}{2})$ for $\rm{^3H}$ and $\rm{^3He}$, respectively.

The one-body and two-body density distributions are defined as
\begin{eqnarray}
\rho_{N}(r)= \frac{1}{M_{N}} \Big\langle \Psi \Big| \sum_{i}^{M_{N}} P_{N}^{(i)} \delta(r-|\boldsymbol{r}_i-\boldsymbol{R}_{cm} |) \Big|\Psi \Big\rangle,
\end{eqnarray}
\begin{eqnarray}
\rho_{NN}(r)= \frac{1}{M_{NN}}  \Big\langle \Psi \Big| \sum_{i<j}^{M_{NN}} P_{NN}^{(ij)} \delta(r-|\boldsymbol{r}_i-\boldsymbol{r}_{j} |) \Big|\Psi \Big\rangle,
\end{eqnarray}
respectively. $P_N^{(i)}=\frac{1}{2}(1\pm \tau_{z,i})$ and $P_{NN}^{(ij)}=\frac{1}{4}(1\pm \tau_{z,i})(1\pm \tau_{z,j})$ are one-body and two-body isospin projection operators, respectively and the subscript $N$ labels proton $p$ and neutron $n$. $M_{N}$ and $M_{NN}$ stand for the number of corresponding nucleon and nucleon-nucleon pair. 
The normalizations are $4\pi\int \rho_{N}(r)r^2dr=1$ and $4\pi\int \rho_{NN}(r)r^2dr=1$.

The basis wave functions of a 3-nucleon system in momentum space are obtained by the Fourier transform of the Gaussian basis functions in coordinate space, $\varphi(\boldsymbol{k})=(\frac{1}{2\pi})^{3/2} \int \phi(\boldsymbol{r}) e^{-i\boldsymbol{k}\cdot \boldsymbol{r} } d \boldsymbol{r}$
and 
$\varphi'(\boldsymbol{K})=(\frac{1}{2\pi})^{3/2} \int \psi(\boldsymbol{R}) e^{-i\boldsymbol{K}\cdot \boldsymbol{R} } d \boldsymbol{R}$.
Then using the 3-body Jacobi coordinates in momentum space,
$\boldsymbol{k}_1=(\boldsymbol{p}_3-\boldsymbol{p}_2)/2$, 
$\boldsymbol{k}_2=(\boldsymbol{p}_1-\boldsymbol{p}_3)/2$,
$\boldsymbol{k}_3=(\boldsymbol{p}_2-\boldsymbol{p}_1)/2$,
$\boldsymbol{K}_1=2(\boldsymbol{p}_1-\frac{1}{2}\boldsymbol{p}_2-\frac{1}{2}\boldsymbol{p}_3)/3$,
$\boldsymbol{K}_2=2(\boldsymbol{p}_2-\frac{1}{2}\boldsymbol{p}_3-\frac{1}{2}\boldsymbol{p}_1)/3$,
and
$\boldsymbol{K}_3=2(\boldsymbol{p}_3-\frac{1}{2}\boldsymbol{p}_1-\frac{1}{2}\boldsymbol{p}_2)/3$,
we obtain the total wave function $\Phi$ of the 3-nucleon system in momentum space,
\begin{eqnarray}
&&\Phi_{TM_T,JM}=\sum_{C=1}^3 \sum_{\alpha}A_{\alpha}
\Big[ [\eta_{\frac{1}{2}}\eta_{\frac{1}{2}}]_t \eta_{\frac{1}{2}} \Big]_{TM_T} \times \notag \\
&& \ \ \ \ \  \Big[ \big[ [\chi_{\frac{1}{2}}\chi_{\frac{1}{2}}]_s \chi_{\frac{1}{2}} \big]_{S} \big[ \varphi_{nl}(\boldsymbol{k}^{(C)} ){\varphi'}_{NL}(\boldsymbol{K}^{(C)}) \big]_I \Big]_{JM},
\end{eqnarray}
where $\boldsymbol{k}$ and $\boldsymbol{K}$ stand for the relative momentum between two nucleons and relative momentum between $NN$ pair and the third nucleon, respectively. The c.o.m. momentum of $NN$ pair $\boldsymbol{Q}=-\boldsymbol{K}$ when we omit the c.o.m. motion of the 3-nucleon system. 

\section{Results}

We firstly calculate the binding energies (B.E.) and the root-mean-square (R.M.S.) radii for $\rm{^3H}$ and $\rm{^3He}$, respectively. The R.M.S. radius $R$ is defined as $R=(\int r^2 \rho(r) r^2 dr / \int \rho(r) r^2 dr )^{1/2} $. 
In the diagonalization of the 3-body Hamiltonian, we use basis functions with $l\leq 2$ and $L\leq 2$, which are enough to make the eigenvalues convergence quickly. 
The comparison between the calculated results and the experimental data is given in Table \ref{ene}. It is clearly shown that the binding energies and the proton R.M.S. radii for the bound states of $\rm{^3H}$ and $\rm{^3He}$ nuclei are both well reproduced. 
We also calculate the expectation values of kinetic energy, each part of potential energies and potential energies in different isospin $t$ and spin $s$ channels. The results, which are listed in Table \ref{expectation}, show that the central potential exists in all $(t,s)$ channels but mainly contributes as attraction in the $(0,1)$ and $(1,0)$ channels. Spin-orbit potential and tensor potential only exist in $s=1$ channels. It should be emphasized that the tensor potential in the $(t,s)=(0,1)$ channel is important and contributes $\sim 55\%$ of total attraction. Repulsive Coulomb potential is considered only between $pp$ in the $t=1$ channels for $\rm{^3He}$. The 3-body interaction serves as an attractive potential for both $\rm{^3H}$ and $\rm{^3He}$ and its contribution is relatively small.

\begin{figure}[b]
\centering
\includegraphics[width=8.6cm,angle=0,clip=true]{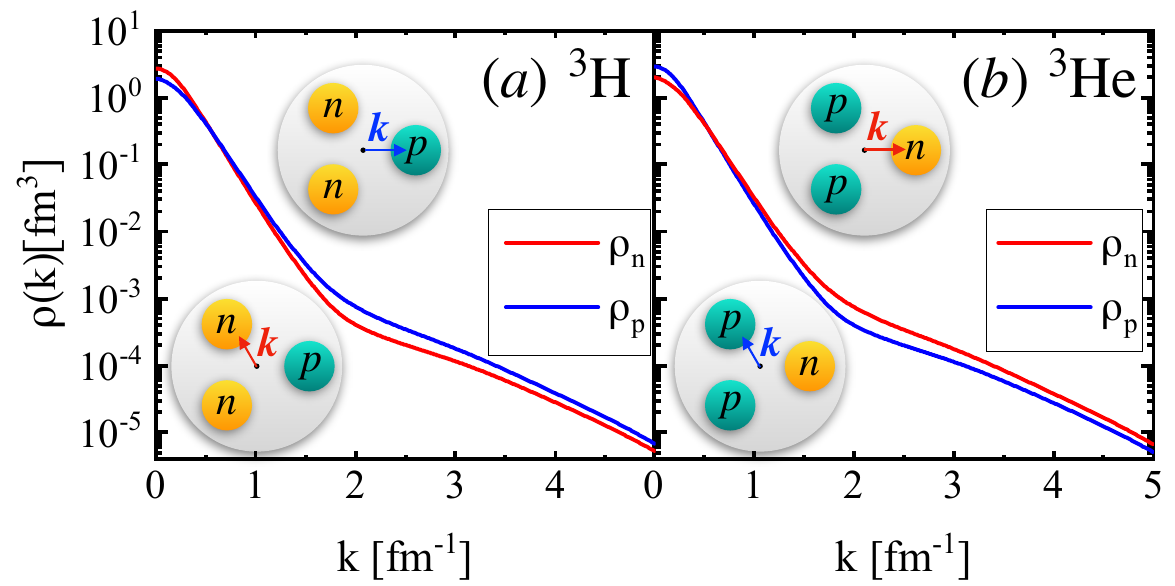}\\
  \caption{One-body proton ($\rho_p$) and neutron ($\rho_n$) momentum distributions for $\rm{^3H}$ and $\rm{^3He}$ as functions of the momentum $k$.}
  \label{1b}
\end{figure}

The one-body momentum distribution is defined as
\begin{eqnarray}
\label{1ppn}
\rho_{N}(k)=\frac{1}{M_{N}}\Big\langle \Phi \Big| \sum_{i}^{M_{N}} P_{N}^{(i)} \delta(k-|\boldsymbol{k}_i|) \Big|\Phi  \Big\rangle,
\end{eqnarray}
with the normalization condition $4\pi\int \rho_{N}(k)k^2dk=1$ and the c.o.m. motion of the 3-nucleon system is omitted. 
In Fig.~\ref{1b}$(a)$ and Fig.~\ref{1b}$(b)$, we display the calculated one-body proton and neutron momentum distributions for $\rm{^3H}$ and $\rm{^3He}$, respectively. One can see that the proton and neutron momentum distributions both have max values at $k=0 \ \rm{fm^{-1}}$, and fall down rapidly in the range of $0 < k< 2.0$ $\rm{fm}^{-1}$. As expected, the high-momentum tail appears with $k>2 \ \rm{fm^{-1}}$, which is attributed to the effect of SRCs between pairs of nucleons. 
But differences are seen between the proton and neutron for $\rm{^3H}$ and $\rm{^3He}$, namely, the minority nucleon (proton for $\rm{^3H}$ and neutron for $\rm{^3He}$) has larger high-momentum tail. 
This is considered to be a natural consequence of the short-range tensor interaction. 
Taking $\rm ^3He$ as an example, the $pn$-SRC generated from tensor interaction populates one proton and one neutron in high-momentum state while the remaining proton (majority nucleon) is in relatively low momentum state.   
Thus the neutron (minority nucleon) has larger high-momentum tail and larger kinetic energy compared with the proton.
This feature also manifests itself in heavy nuclei such as $\rm ^{27}Al$, $\rm ^{56}Fe$ and $\rm ^{208}Pb$. The average proton kinetic energy in these nuclei is found to be larger than that of the neutron one in a $pn$-dominance toy model \cite{Hen:2014nza}.

\begin{figure}[t]
\centering
\includegraphics[width=8.6cm,angle=0,clip=true]{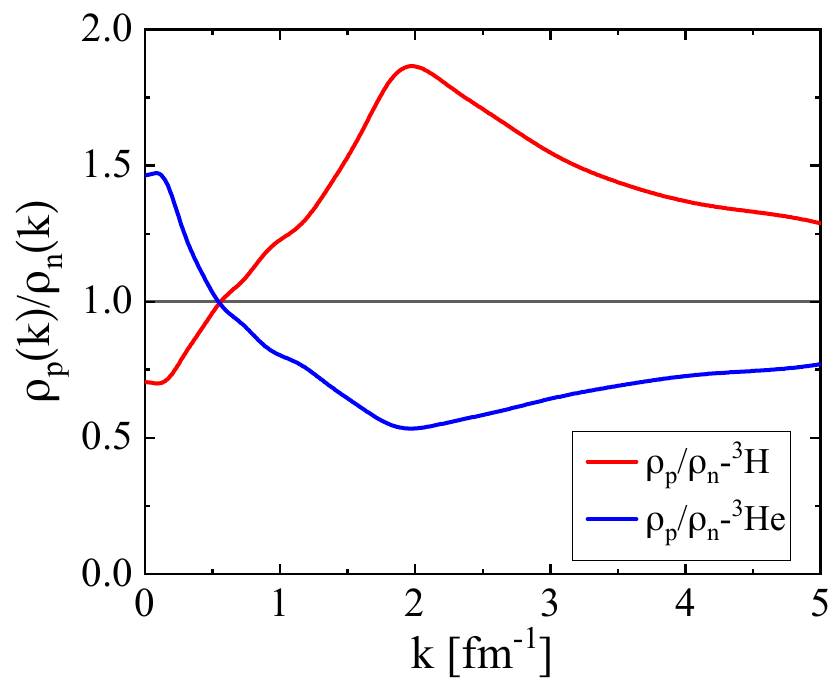}\\
  \caption{Ratios of proton to neutron ($\rho_p/\rho_n$) for $\rm{^3H}$ and $\rm{^3He}$ as functions of the momentum $k$.}
  \label{ratio-1b}
\end{figure}

Fig.~\ref{ratio-1b} shows the ratios of proton momentum distribution to neutron one ($\rho_p/\rho_n$) for $\rm{^3H}$ (red curve) and $\rm{^3He}$ (blue curve). Two curves are roughly symmetric about the horizontal line $\rho_p(k)/\rho_n(k)=1$. The behavior of the ratio is determined mainly by the competition between the tensor interaction and the repulsive hard-core. Taking the red curve for $\rm{^3H}$ as an example, the ratio of minority nucleon to majority nucleon keeps increasing in the range of $0<k<2.0$ $\rm{fm}^{-1}$. This is expected because the tensor interaction plays a more and more important role with the increasing of $k$. The decreasing of the ratio beyond $k=2.0$ $\rm{fm^{-1}}$ is because the short-range repulsive hard-core starts to contribute largely and reduces the dominance of tensor interaction. Note that the short-range repulsive hard-core exists in all $NN$ channels including $nn$ and $pp$ channels. For very large $k$, the ratio of $\rho_p/\rho_n$ is expected to become smaller. 
At $k=5.0 \ \rm{fm^{-1}}$, this ratio reduces to approximately $\rho_p/\rho_n=1.25$, indicating that the tensor interaction still contributes but in less dominance.
The blue line for $\rm{^3He}$ has similar behavior except that it is shown in the ratio of majority nucleon to minority one. We do not repeat the discussion here.

\begin{center}
\renewcommand\arraystretch{1.1} 
\begin{table}[t]
\caption{The calculated $\rm{^3H}$ and $\rm{^3He}$ binding energies (B.E.) and root-mean-square (R.M.S.) radii using AV8' interaction (Cal.(AV8')) and AV8' and $NNN$ 3-body interaction (Cal.(AV8'+3NI)), compared with the experimental values (Exp.). }
\label{ene}
\begin{tabular}{p{0.8cm}<{\centering}p{1.5cm}<{\centering} p{1.8cm}<{\centering} p{2.2cm}<{\centering} p{1.5cm}<{\centering} }
\hline\hline
 &  &Cal.(AV8')&Cal.(AV8'+3NI)&Exp. \\
\hline
$\rm{^3H}$ &B.E.(MeV)    & -7.77 & -8.44 & -8.48 \\
 &$R_p$(fm)    & 1.637 & 1.597 & 1.59 \\
 &$R_n$(fm)    & 1.790 & 1.740 &  \\
 &$R_{pn}$(fm) & 2.922 & 2.846 &  \\
 &$R_{nn}$(fm) & 3.189 & 3.094 &  \\
\hline
$\rm{^3He}$ &B.E.(MeV)    & -7.11 & -7.76 & -7.72 \\
 &$R_p$(fm)    & 1.824 & 1.770 & 1.76 \\
 &$R_n$(fm)    & 1.660 & 1.617 &  \\
 &$R_{pn}$(fm) & 2.967 & 2.886 &  \\
 &$R_{pp}$(fm) & 3.256 & 3.152 &  \\
\hline\hline
\end{tabular}
\end{table}
\end{center}

\begin{center}
\renewcommand\arraystretch{1.1} 
\begin{table}[t]
\caption{The expectation values of kinetic energy $K$ and potential energies of central $\langle V^{Cen} \rangle$, spin-orbit $\langle V^{LS} \rangle$, tensor $\langle V^{Ten} \rangle$, Coulomb $\langle V^{Coul} \rangle$ and $NNN$ 3-body interaction $\langle V^{NNN} \rangle$ in different isospin $t$ and spin $s$ channels for $\rm{^3H}$ and $\rm{^3He}$ (Unit:MeV). }
\label{expectation}
\begin{tabular}{p{0.5cm}<{\centering}p{0.3cm}<{\centering} p{0.3cm}<{\centering} p{1.0cm}<{\centering} p{1.0cm}<{\centering} p{1.0cm}<{\centering} p{1.0cm}<{\centering} p{1.0cm}<{\centering}p{1.0cm}<{\centering} }
\hline\hline
  & \multirow{2}{*}{$t$} 
  & \multirow{2}{*}{$s$} 
  & \multirow{2}{*}{$K$} 
  & 
\multicolumn{3}{c}{AV8'}
  & \multirow{2}{*}{$\langle V^{Coul} \rangle$} 
  & \multirow{2}{*}{$\langle V^{NNN} \rangle$} \\
  \cline{5-7}
  		   &  &  &   & $\langle V^{Cen} \rangle$   & $\langle V^{LS} \rangle$  & $\langle V^{Ten} \rangle$  &  &   \\
\hline
\multirow{5}{*}{$\rm{^3H}$}& 0 & 0 &   & 0.02   & 0     & 0      & 0 &    \\
		  & 0 & 1 &   & -8.72  & -1.97 & -31.50 & 0 &   \\
		  & 1 & 0 &   & -14.74 & 0 	 & 0 	  & 0 &   \\
		  & 1 & 1 &   & 0.19   & -0.10 & -0.24  & 0 &   \\
\cline{2-9}
          &  \multicolumn{2}{c}{sum} & 49.54 &  -23.25  &   -2.07   &   -31.74    & 0  & -0.92  \\
		  
\hline
\multirow{5}{*}{$\rm{^3He}$}& 0 & 0 &   & 0.02   & 0	 & 0	  & 0 &   \\
		   & 0 & 1 &   & -8.63  & -1.95 & -31.16 & 0 &   \\
		   & 1 & 0 &   & -14.36 & 0	 & 0	  & 0.61 &   \\
		   & 1 & 1 &   & 0.19   & -0.10 & -0.23  & 0.06 &   \\
    \cline{2-9}
           &  \multicolumn{2}{c}{sum} & 48.69 &  -22.78   & -2.05	 & 	-31.39  & 0.67 & -0.91 \\
\hline\hline
\end{tabular}
\end{table}
\end{center}


\begin{figure}[b]
\centering
\includegraphics[width=8.6cm,angle=0,clip=true]{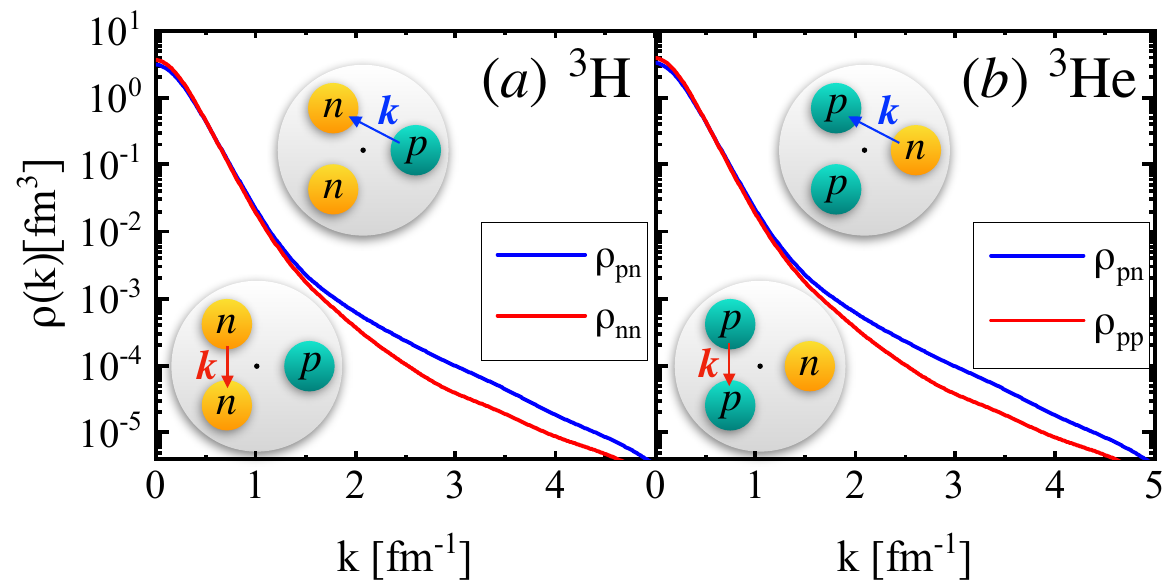}\\
  \caption{Two-body proton-neutron ($\rho_{pn}$) and neutron-neutron ($\rho_{nn}$) momentum distributions for $\rm{^3H}$ and $\rm{^3He}$ as functions of the relative momentum $k$.}
  \label{2b}
\end{figure}

The two-body momentum distribution $\rho_{NN}$ is a function of $NN$ relative momentum $k$ after integrating over all values of of c.o.m. momentum of $NN$ pairs $\boldsymbol{Q}$,
\begin{eqnarray}
\label{2ppn}
\rho_{NN}(k)=\frac{1}{M_{NN}} \Big\langle \Phi \Big| \sum_{i<j}^{M_{NN}} P_{NN}^{(ij)} \delta(k-|\boldsymbol{k}_i-\boldsymbol{k}_{j} |) \Big|\Phi  \Big\rangle,
\end{eqnarray}
with the normalization $4\pi\int \rho_{NN}(k)k^2dk=1$.
We display the calculated two-body momentum distributions of different $NN$ pairs for $\rm{^3H}$ and $\rm{^3He}$ in Fig.~\ref{2b}$(a)$ and Fig.~\ref{2b}$(b)$, respectively, with $k$ ranging from $0$ to $5.0 \ \rm{fm^{-1}}$. In general, the behavior of two-body momentum distributions is similar to that of one-body ones. When $k>2 \ \rm{fm}^{-1}$, the $pn$ pair in $\rm{^3H}$ shows a large high-momentum tail while that in the $nn$ pair is much smaller. Similar to the case of $\rm{^3H}$, the high-momentum tail appears in the $pn$ pairs rather than in the $pp$ pair for $\rm{^3He}$.

\begin{figure}[h]
\centering
\includegraphics[width=8.6cm,angle=0,clip=true]{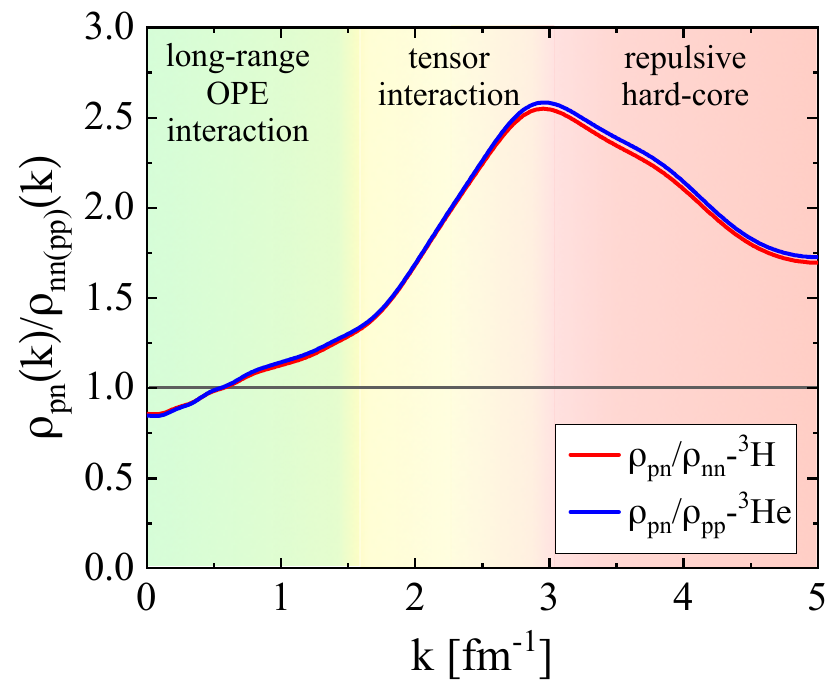}\\
  \caption{Ratios of $pn$ to $nn$ pairs ($\rho_{pn}/\rho_{nn}$) for $\rm{^3H}$ and $pn$ to $pp$ pairs ($\rho_{pn}/\rho_{pp}$) for $\rm{^3He}$ as function of the relative momentum $k$.}
  \label{ratio-2b}
\end{figure}

The ratios of $pn$ to $pp$($nn$) pairs as function of $k$ are shown in Fig.~\ref{ratio-2b} (red curve for $\rm{^3H}$ and blue curve for $\rm{^3He}$), which can be approximately divided into three regions.
The first region ($0<k<1.5 \ {\rm fm^{-1}}$) is considered to be dominated by the long-range one-pion-exchange potential and the variation of $pn$/$pp(nn)$ ratio is rather smooth. The rapid changing of slope in the second region ($1.5 \ {\rm fm^{-1}}<k<3.0 \ \rm{fm^{-1}}$) is because the $pn$ pair correlation has a strong dominance compared with the $pp$ and $nn$ correlations, due to the strong tensor interaction of the $pn$ pair. This is consistent with the latest results of $ab$ $initio$ variational Monte Carlo (VMC) calculations \cite{Piarulli:2022ulk}. As discussed above, the decreasing of the ratios in the third region ($3.0 \ {\rm fm^{-1}}<k<5.0 \ \rm{fm^{-1}}$) is very likely because that the repulsive hard-core from $NN$ scalar interaction becomes dominating. Since the strong repulsive core exists in both $pn$ pair and $pp(nn)$ pair, the $pn$ pair correlation becomes less dominating and the ratio of $pn$ to $pp$($nn$) pairs becomes smaller.
Note that the red and blue curves almost coincide with each other but small difference is found between two curves with $k>2.5 \ \rm{fm}^{-1}$, {\it i.e.} the ratio of $pn/pp$ is larger than that of $pn/nn$. This can be explained by the Coulomb interaction between $pp$, which is repulsive and makes the $pp$ correlation weaker than the $nn$ one, leading to the small difference between $\rm ^3H$ and $\rm ^3He$.

\begin{center}
\renewcommand\arraystretch{1.0} 
\begin{table}[h]
\caption{The calculated non-SRC and SRC $pn$ and $nn$($pp$) pairs percentage and ratios of $pn$ SRC pairs to $nn(pp)$ SRC ones. 
For non-SRC pairs and SRC pairs, we define that $P_{NN}=\int_0^{k_F}\rho_{NN}(k)k^2dk/\int_0^{\infty}\rho_{NN}(k)k^2dk$ and $P_{NN}=\int_{k_F}^{k_{max}}\rho_{NN}(k)k^2dk/\int_0^{\infty}\rho_{NN}(k)k^2dk$, respectively, where $k_{max}=5.0 \ \rm{fm}^{-1}$.}
\label{int}
\begin{tabular}{p{0.6cm}<{\centering}p{1.2cm}<{\centering} p{1.2cm}<{\centering} p{0.1cm}<{\centering}p{1.4cm}<{\centering} p{1.4cm}<{\centering} p{1.8cm}<{\centering} }
\hline\hline
 & \multicolumn{2}{c}{non-SRC pairs} & & \multicolumn{2}{c}{SRC pairs} & SRC  \\
 \cline{2-3}
 \cline{5-6}
 & $P_{pn}(\%)$ & $P_{nn(pp)}(\%)$ && $P_{pn}(\%)$ & $P_{nn(pp)}(\%)$ & $P_{pn}/P_{nn(pp)}$ \\
\hline
$\rm{^3H}$   & 93.57 & 96.02 && 6.43 & 3.98 & 1.62 \\
$\rm{^3He}$  & 93.67 & 96.11 && 6.33 & 3.89 & 1.63 \\
\hline\hline
\end{tabular}
\end{table}
\end{center}

Finally we discuss the enhancement of $pn$-SRC pairs to $nn(pp)$-SRC ones in 3-nucleon systems and make comparison between theory and experimental data. Here the SRC pair is defined as $NN$ pair with relative momentum $k$ larger than the Fermi momentum in saturated nuclear matter $k_F=1.33 \ \rm fm^{-1}$. 
The percentages of non-SRC pairs with relative momentum below the Fermi momentum $0<k< k_F$ and SRC pairs with relative momentum $ k_F<k<5.0 \ \rm{fm}^{-1}$ are calculated for $\rm{^3H}$ or $\rm{^3He}$, respectively. 
The results are listed in Table \ref{int}. It is clearly seen that a small percentage of $pn$ and $nn(pp)$-SRC pairs is found in both nuclei (3.89\%-6.43\%). 
For these SRC pairs, the ratios of $P_{pn}(\%)$ to $P_{nn(pp)}(\%)$ are 1.62 and 1.63 for $\rm{^3H}$ and $\rm{^3He}$, respectively. 
This means that the percentages of $pn$-SRC pairs are significantly enhanced compared with those of $nn(pp)$-SRC ones in both nuclei. 
The enhancement of $pn$-SRC pairs is consistent with the experimental data \cite{Li:2022fhh}.  
Note that an enhancement factor of $2.17^{+0.25}_{-0.20}$ is extracted from the experiment data with c.o.m. momentum of $NN$ pair $\boldsymbol{Q}\simeq 0$.  
For $\boldsymbol{Q}>0$,  higher partial wave components are expected to be involved, leading to a higher average relative momentum \cite{Wiringa:2008dn}.  
Thus, a relatively smaller enhancement factor can be obtained by summing over all values of $\boldsymbol{Q}$. It should be interesting to compare the theoretical results with the experimental data for $\boldsymbol{Q}>0$. One would expect the ratio of $pn$-SRC pairs to $nn(pp)$ ones to be smaller and very likely to be more consistent with our theoretical prediction. 

\section{Summary}

We have performed microscopic calculations of the one-, and two-nucleon momentum distributions and the $pn/nn(pp)$ SRC ratios for mirror nuclei $\rm{^3H}$ and $\rm{^3He}$. We show that the $pn$-SRCs are enhanced compared with the $nn(pp)$-SRCs, which is consistent with the recent experimental data. We also show that the tensor-force-induced SRC competes strongly with the hard-core-induced SRC beyond the Fermi momentum. The tensor SRC pairs dominate in the inter-medium region of $1.5 \ {\rm fm^{-1}}<k<3.0 \ \rm{fm^{-1}}$ while the hard-core SRC ones in higher momentum region $k>3.0 \ \rm{fm^{-1}}$. The present microscopic GEM calculations can possibly be extended to heavier systems in which the percentage of $pn$-SRCs is expected to be further enhanced. A comparison of 3-nucleon systems and heavier ones should be helpful to better understand the short-distance part of nuclear force and its isospin-dependence.

\begin{acknowledgments}
The authors would like to thank Zhihong Ye, Mengjiao Lyu, and Emiko Hiyama for the helpful discussions.
The work is supported by the National Natural Science Foundation of China (Grant No. 12275129) and the Fundamental Research Funds for the Central Universities (Grant No. 020414380209).
\end{acknowledgments}


\begin{thebibliography}{99}

\bibitem{Hen:2016kwk}
O.~Hen, G.~A.~Miller, E.~Piasetzky and L.~B.~Weinstein,
Rev. Mod. Phys. \textbf{89}, 045002 (2017).




\bibitem{Reid:1968sq}
R.~V.~Reid, Jr.,
Annals Phys. \textbf{50}, 411 (1968).

\bibitem{Lacombe:1980dr}
M.~Lacombe, B.~Loiseau, J.~M.~Richard, R.~Vinh Mau, J.~Cote, P.~Pires and R.~De Tourreil,
Phys. Rev. C \textbf{21}, 861(1980).

\bibitem{Pudliner:1997ck}
B.~S.~Pudliner, V.~R.~Pandharipande, J.~Carlson, S.~C.~Pieper and R.~B.~Wiringa,
Phys. Rev. C \textbf{56}, 1720 (1997).

\bibitem{Wiringa:1984tg}
R.~B.~Wiringa, R.~A.~Smith and T.~L.~Ainsworth,
Phys. Rev. C \textbf{29}, 1207 (1984).

\bibitem{Wiringa:1994wb}
R.~B.~Wiringa, V.~G.~J.~Stoks and R.~Schiavilla,
Phys. Rev. C \textbf{51}, 38 (1995).







\bibitem{Co:1994bzw}
G.~Co', A.~Fabrocini and S.~Fantoni,
Nucl. Phys. A \textbf{568}, 73(1994).

\bibitem{Fabrocini:1999mz}
A.~Fabrocini, F.~Arias de Saavedra and G.~Co',
Phys. Rev. C \textbf{61}, 044302 (2000).

\bibitem{Bisconti:2007vu}
C.~Bisconti, F.~Arias de Saavedra and G.~Co',
Phys. Rev. C \textbf{75}, 054302 (2007).

\bibitem{Ryckebusch:2014ann}
J.~Ryckebusch, W.~Cosyn and M.~Vanhalst,
J. Phys. G \textbf{42}, 055104 (2015).


\bibitem{Dickhoff:2004xx}
W.~H.~Dickhoff and C.~Barbieri,
Prog. Part. Nucl. Phys. \textbf{52}, 377(2004).

\bibitem{Rios:2013zqa}
A.~Rios, A.~Polls and W.~H.~Dickhoff,
Phys. Rev. C \textbf{89}, 044303 (2014).



\bibitem{Alvioli:2007zz}
M.~Alvioli, C.~Ciofi degli Atti and H.~Morita,
Phys. Rev. Lett. \textbf{100}, 162503 (2008).

\bibitem{Alvioli:2012qa}
M.~Alvioli, C.~Ciofi degli Atti, L.~P.~Kaptari, C.~B.~Mezzetti and H.~Morita,
Phys. Rev. C \textbf{87},  034603 (2013).

\bibitem{Alvioli:2011aa}
M.~Alvioli, C.~Ciofi degli Atti, L.~P.~Kaptari, C.~B.~Mezzetti, H.~Morita and S.~Scopetta,
Phys. Rev. C \textbf{85}, 021001 (2012).

\bibitem{Alvioli:2016wwp}
M.~Alvioli, C.~Ciofi degli Atti and H.~Morita,
Phys. Rev. C \textbf{94},  044309 (2016).

\bibitem{Lyu:2019bxr}
M.~Lyu, T.~Myo, H.~Toki, H.~Horiuchi, C.~Xu and N.~Wan,
Phys. Lett. B \textbf{805}, 135421 (2020).

\bibitem{Weiss:2016obx}
R.~Weiss, R.~Cruz-Torres, N.~Barnea, E.~Piasetzky and O.~Hen,
Phys. Lett. B \textbf{780}, 211-215 (2018).
\bibitem{Schiavilla:2006xx}
R.~Schiavilla, R.~B.~Wiringa, S.~C.~Pieper and J.~Carlson,
Phys. Rev. Lett. \textbf{98}, 132501 (2007).

\bibitem{Wiringa:2008dn}
R.~B.~Wiringa, R.~Schiavilla, S.~C.~Pieper and J.~Carlson,
Phys. Rev. C \textbf{78}, 021001 (2008).

\bibitem{Wiringa:2013ala}
R.~B.~Wiringa, R.~Schiavilla, S.~C.~Pieper and J.~Carlson,
Phys. Rev. C \textbf{89},  024305 (2014).

\bibitem{Piarulli:2022ulk}
M.~Piarulli, S.~Pastore, R.~B.~Wiringa, S.~Brusilow and R.~Lim,
Phys. Rev. C \textbf{107},  014314 (2023).

\bibitem{Carlson:2014vla}
J.~Carlson, S.~Gandolfi, F.~Pederiva, S.~C.~Pieper, R.~Schiavilla, K.~E.~Schmidt and R.~B.~Wiringa,
Rev. Mod. Phys. \textbf{87}, 1067 (2015).








\bibitem{Tang:2002ww}
A.~Tang, J.~W.~Watson, J.~L.~S.~Aclander, J.~Alster, G.~Asryan, Y.~Averichev, D.~Barton, V.~Baturin, N.~Bukhtoyarova and A.~Carroll, \textit{et al.}
Phys. Rev. Lett. \textbf{90}, 042301 (2003).

\bibitem{CLAS:2005ola}
K.~S.~Egiyan \textit{et al.} [CLAS],
Phys. Rev. Lett. \textbf{96}, 082501 (2006).

\bibitem{Piasetzky:2006ai}
E.~Piasetzky, M.~Sargsian, L.~Frankfurt, M.~Strikman and J.~W.~Watson,
Phys. Rev. Lett. \textbf{97}, 162504 (2006).


\bibitem{JeffersonLabHallA:2007lly}
R.~Shneor \textit{et al.} ,
Phys. Rev. Lett. \textbf{99}, 072501 (2007).

\bibitem{Subedi:2008zz}
R.~Subedi, R.~Shneor, P.~Monaghan, B.~D.~Anderson, K.~Aniol, J.~Annand, J.~Arrington, H.~Benaoum, W.~Bertozzi and F.~Benmokhtar, \textit{et al.}
Science \textbf{320}, 1476 (2008).

\bibitem{CLAS:2010yvl}
H.~Baghdasaryan \textit{et al.} [CLAS],
Phys. Rev. Lett. \textbf{105}, 222501 (2010).

\bibitem{Fomin:2011ng}
N.~Fomin, J.~Arrington, R.~Asaturyan, F.~Benmokhtar, W.~Boeglin, P.~Bosted, A.~Bruell, M.~H.~S.~Bukhari, M.~E.~Christy and E.~Chudakov, \textit{et al.}
Phys. Rev. Lett. \textbf{108}, 092502 (2012).

\bibitem{Hen:2014nza}
O.~Hen, M.~Sargsian, L.~B.~Weinstein, E.~Piasetzky, H.~Hakobyan, D.~W.~Higinbotham, M.~Braverman, W.~K.~Brooks, S.~Gilad and K.~P.~Adhikari, \textit{et al.}
Science \textbf{346}, 614 (2014).

\bibitem{LabHallA:2014wqo}
I.~Korover \textit{et al.},
Phys. Rev. Lett. \textbf{113}, 022501 (2014).


\bibitem{CLAS:2018xvc}
M.~Duer \textit{et al.} [CLAS],
Phys. Rev. Lett. \textbf{122},  172502 (2019).

\bibitem{CLAS:2018yvt}
M.~Duer \textit{et al.} [CLAS],
Nature \textbf{560}, 617 (2018).




\bibitem{CiofidegliAtti:2015lcu}
C.~Ciofi degli Atti,
Phys. Rep. \textbf{590}, 1(2015).

\bibitem{Frankfurt:1993sp}
L.~L.~Frankfurt, M.~I.~Strikman, D.~B.~Day and M.~Sargsian,
Phys. Rev. C \textbf{48}, 2451(1993).

\bibitem{Frankfurt:2008zv}
L.~Frankfurt, M.~Sargsian and M.~Strikman,
Int. J. Mod. Phys. A \textbf{23}, 2991(2008).



\bibitem{Arrington:2011xs}
J.~Arrington, D.~W.~Higinbotham, G.~Rosner and M.~Sargsian,
Prog. Part. Nucl. Phys. \textbf{67}, 898 (2012).










\bibitem{Li:2022fhh}
S.~Li, R.~Cruz-Torres, N.~Santiesteban, Z.~H.~Ye, D.~Abrams, S.~Alsalmi, D.~Androic, K.~Aniol, J.~Arrington and T.~Averett, \textit{et al.}
Nature \textbf{609},  41(2022).



\bibitem{Hiyama:2004nf}
E.~Hiyama, B.~F.~Gibson and M.~Kamimura,
Phys. Rev. C \textbf{70}, 031001 (2004).


\bibitem{Hiyama:2003cu}
E.~Hiyama, Y.~Kino and M.~Kamimura,
Prog. Part. Nucl. Phys. \textbf{51}, 223(2003).

\bibitem{Hiyama:2022jqh}
E.~Hiyama, M.~Isaka, T.~Doi and T.~Hatsuda,
Phys. Rev. C \textbf{106}, 064318 (2022).





\bibitem{Hiyama:2022loc}
E.~Hiyama, R.~Lazauskas, J.~Carbonell and T.~Frederico,
Phys. Rev. C \textbf{106}, 064001 (2022).


\bibitem{Hiyama:2018ukv}
E.~Hiyama, A.~Hosaka, M.~Oka and J.~M.~Richard,
Phys. Rev. C \textbf{98}, 045208 (2018).

\bibitem{Meng:2019fan}
Q.~Meng, E.~Hiyama, K.~U.~Can, P.~Gubler, M.~Oka, A.~Hosaka and H.~Zong,
Phys. Lett. B \textbf{798}, 135028 (2019).


\bibitem{Meng:2020knc}
Q.~Meng, E.~Hiyama, A.~Hosaka, M.~Oka, P.~Gubler, K.~U.~Can, T.~T.~Takahashi and H.~S.~Zong,
Phys. Lett. B \textbf{814}, 136095 (2021).


\bibitem{Meng:2021yjr}
Q.~Meng, M.~Harada, E.~Hiyama, A.~Hosaka and M.~Oka,
Phys. Lett. B \textbf{824}, 136800 (2022).



\bibitem{Xu:2010xh}
C.~Xu and B.~A.~Li,
Phys. Rev. C \textbf{81}, 044603 (2010).

\bibitem{Xu:2009bb}
C.~Xu and B.~A.~Li,
Phys. Rev. C \textbf{81}, 064612 (2010).

\bibitem{Xu:2012hf}
C.~Xu, A.~Li and B.~A.~Li,
J. Phys. Conf. Ser. \textbf{420}, 012090 (2013).

\bibitem{Zhang:2014bna}
X.~Zhang, C.~Xu and Z.~Ren,
Eur. Phys. J. A \textbf{50}, 113 (2014).

\bibitem{Carbone:2011wk}
A.~Carbone, A.~Polls and A.~Rios,
EPL \textbf{97}, 22001 (2012).

\bibitem{Vidana:2011ap}
I.~Vidana, A.~Polls and C.~Providencia,
Phys. Rev. C \textbf{84}, 062801 (2011).

\bibitem{Li:2019xxz}
B.~A.~Li, P.~G.~Krastev, D.~H.~Wen and N.~B.~Zhang,
Eur. Phys. J. A \textbf{55}, 117 (2019).


\end{thebibliography}
\end{document}